\documentstyle[12pt]{article}
\newcommand{\beq}{\begin{equation}}
\newcommand{\eeq}{\end{equation}}
\newcommand{\bdm}{\begin{displaymath}}
\newcommand{\edm}{\end{displaymath}}
\newcommand{\bea}{\begin{eqnarray}}
\newcommand{\eea}{\end{eqnarray}}

\newcommand{\th}{\theta}
\newcommand{\thb}{\bar{\theta}}
\newcommand{\ab}{\bar{a}}
\newcommand{\bb}{\bar{b}}
\newcommand{\cb}{\bar{c}}

\begin{document}
\setcounter{page}{0}
\topmargin 0pt
\oddsidemargin 5mm
\renewcommand{\thefootnote}{\fnsymbol{footnote}}
\newpage
\setcounter{page}{0}
\begin{titlepage}
\begin{flushright}
USP-IFQSC/TH/93-06
\end{flushright}
\vspace{0.5cm}
\begin{center}
{\large {\bf Factorized Scattering in the Presence of  Reflecting
Boundaries}} \\
\vspace{1.8cm}
{\large Andreas Fring}\footnote{ Supported by FAPESP - Brasil}
{\large and Roland K\"oberle\footnote{ Supported in part by CNPq-Brasil.}}
\footnote{ FRING@BR.ANSP.USP.IFQSC  and  ROLAND@IFQSC.ANSP.BR } \\
\vspace{0.5cm}
{\em Universidade de S\~ao Paulo, \\
Caixa Postal 369, CEP 13560 S\~ao Carlos-SP, Brasil\\}
\vspace{3cm}
\renewcommand{\thefootnote}{\arabic{footnote}}
\setcounter{footnote}{0}
\begin{abstract}
{ We formulate a general set of consistency requirements, which are
expected to be satisfied by the scattering matrices in the presence of
reflecting boundaries. In particular we derive an equivalent to the
boostrap equation involving the W-matrix, which encodes the reflection
of a particle off a wall. This set of equations is
sufficient to derive explicit formulas for $W$, which we illustrate in the case
   of some particular affine Toda field theories.}
\end{abstract}
\vspace{.3cm}
\centerline{February 1993}
 \end{center}
\end{titlepage}
\newpage
\section{Introduction}

The common procedure to treat the scattering of particles is to  work in
infinit
   ely extended space-time. Yet restricting the space volume to finite size may
   reveal interesting information, which is not observable in the infinite
volum
   e limit. In completely integrable models the modifications arising from the
p
   resence of boundaries can be computed exactly.
 We therefore direct our interest to integrable models on a finite line
delimite
   d by perfectly reflecting mirrors. The central object is the $S$-matrix,
whic
   h is factorized into 2-body $S$-matrices in this case. They have to satisfy
Y
   ang-Baxter\cite{YB,ZZ} equations, which provide nontrivial constraints for
no
   n-diagonal matrices. In the presence of reflecting boundaries one obtains
similar factorization equations including the {\em wall matrix} $W$, which
descr
   ibes the scattering of a particle off the wall \cite{Ch1,Ch2,Sk}.

The main object of the present paper is to study scattering described by
diagona
   l $S$-matrices. In this case the nontrivial constraints  result from the
boot
   strap hypothesis, which we formulate for the situation with finite space
volu
   me. Whilst
in the situation without boundaries the ensueing consistency equations allow us
to determine explicitly the $S$-matrix \cite{AFZ,KS}, in the present situation
t
   hey will enable us to compute the $W$-matrix.

The layout of this article is as follows. Firstly we extend the Zamolodchikov
al
   gebra including the wall matrix $W$ to take boundaries into account. In
section 3 we employ it to derive the factorization equations in the presence
of reflecting boundaries and in section 4 we formulate our central equations
(\ref{eq: inhomboot}), the {\em wall bootstrap } equations. In section 5
we apply this framework to the 1-particle  Bullough-Dodd model
($ A_2^{(2)}$-affine Toda theory) and to several 2-particle affine Toda systems
($ A_2^{(1)}$ and $ A_4^{(2)}$). Finally we state our conclusions.

\section{Zamolodchikov algebra}

Factorized $S$-matrices, describing one-dimensional scattering, have to satisfy
   certain consistency conditions, which in general provide powerful tools for
their explicit construction. These axioms can be extracted most easily as
associ
   ativity conditions   of the well known  {\em Zamolodchikov algebra}
\cite{ZZ}
\bea
Z_a(\th_a) Z_b(\th_b) &=& S_{ab}^{\ab \bb}(\th_{ab}) \,
Z_{\bb}(\th_b) Z_{\ab} (\th_b)   \label{Zamalga}     \\
Z_a^{\dag}(\th_a) Z_b^{\dag}(\th_b) &=& S_{ab}^{\ab \bb}(\th_{ab})\,
Z_{\bb}^{\dag}(\th_b) Z_{\ab}^{\dag}(\th_a)  \label{Zamalgb} \\
Z_a(\th_a) Z_b^{\dag}(\th_b) &=& S_{ab}^{\ab \bb}(\th_{ba}) \,
Z_{\bb}^{\dag}(\th_b) Z_{\ab}(\th_a) + 2\pi \;\delta_{ab} \;\delta({\th_{ab}})
\;\; , \label{Zamalgc}
\eea
where each of the operators $Z_a$ is associated with a particle ``a" and
$S$ denotes the unitary and crossing invariant two particle scattering
matrix which satisfy the Yang-Baxter-(\ref{eq: yangbax}) and bootstrap
equation (\ref{eq: boots}). Their dependence
on the momenta is parameterized usually by the rapidities $\th_a$ , i.e.
$p_a = m (\cosh \th_a,\sinh \th_a)$, having the advantage that the branch cuts
on the real axis in the complex plane of the Mandelstam variable unfold.
Relativistic invariance demands that the scattering matrix depends only
on the rapidity difference, which we denote $\th_{ab} := \th_a - \th_b$.
\par
In the present case this operator algebra has  to be extended in order to
include the presence of a wall.
When a particle scatters off the wall, it reverses its momentum and possibly
cha
   nges its quantum numbers. If $Z_a(\th)$ represents particle $a$ and $Z_w(0)$
   represents the wall in Zamolodchikov's algebra, this process is encoded in
th
   e following relation
\beq
\label{w1}
        Z_a(\th) Z_w(0) = \sum_{\ab} W_a^{\ab}(\th) Z_{\ab}(\thb) Z_w(0),
\eeq
where $\thb=-\th$ and the matrix $W_{a}^{\ab}(\th)$ describes the scattering by
   the wall. Notice that we do not interchange the order of $Z_a$ and $Z_w$ as
i
   n (\ref{Zamalga})-(\ref{Zamalgc}),
such that the $W$-matrix is not the result of a braiding like the scattering
matrix. From its definition, $W_{a}^{\ab}(\th)$ has to satisfy the usual
unitarity condition
\beq
\label{unitarity}
    \sum_{\ab}    W_a^{\ab}(\th)W_{\ab}^{ a'}(-\th)=\delta_a^{a'}.
\eeq
The algebra, involving $Z_a$'s and $Z_a^{\dag}$ only, is now very similar to
the
    usual case (\ref{Zamalga}) - (\ref{Zamalgc}), except that in the process
$a+
   b\rightarrow c+d$, we have to distinguish three different situations, in
which the braiding of two operators might produce:
\begin{enumerate}
\item
     $^-S^{cd}_{ab}(\th_a,\th_b)$: describing scattering before any particle
has
    hit the wall;\item
    $^0S^{cd}_{ab}(\th_a,\th_b)$: describing scattering after one particle has
h
   it the wall;
\item
     $^+S^{cd}_{ab}(\th_a,\th_b)$:  describing scattering after both particles
     have hit the wall.
\end{enumerate}
 In $^0S$, it is the particle with negative rapidity, which has scattered off
 the wall. Notice that the wall breaks translational invariance, so that the
$S$
   -matrices will not depend only on the difference of rapidities. In
particular $^0S$ is in general a function of the sum of the rapidities
$\tilde{
   \th}_{ab} :=\th_a+\th_b$.

\section{Factorization  equations}
We now use the associativity of the previous algebra to derive consistency
condi
   tions.
 Let us therefore consider the scattering of particles labelled by  quantum
numb
   ers $a,b\ldots$, with rapidities $\th_a,\th_b\ldots$ in the presence of a
ref
   lecting wall, which we locate for convenience at rapidity $\th=0$. We start
f
   rom a state with $\th_a>\th_b$, then
\bea
Z_a(\th_a) Z_b(\th_b)  \!\!\!\!\!\!\! & &  \!\!\!\!\!\!\! Z_w(0)   =
      ^-\!S_{ab}^{a_1b_1}(\th_{ab}) \;Z_{b_1}(\th_b) Z_{a_1}(\th_a) Z_w(0)
                                                  \nonumber\\
 \!\!\!&=& \!\!\!    ^-S_{ab}^{a_1b_1}(\th_{ab})W_{a_1}^{\ab_1}(\th_{a_1})
      \; Z_{b_1}(\th_b) Z_{\ab_1}(\thb_a) Z_w(0)  \\
 \!\!\!&=& \!\!\!  ^-S_{ab}^{a_1b_1}(\tilde{\th}_{ab})W_{a_1}^{\ab_1}(\th_a)
      ^0S_{b_1\ab_1}^{b_2\ab_2}(\tilde{\th}_{b_1 a_1})
      \; Z_{\ab_2}(\thb_a)Z_{b_2}(\th_b) Z_w(0)   \nonumber\\
 \!\!\!\!&=& \!\!\! \!  ^-S_{ab}^{a_1b_1}(\th_{ab})W_{a_1}^{\ab_1}(\th_a)
    ^0S_{b_1\ab_1}^{b_2\ab_2}(\tilde{\th}_{b_1 a_1}) W_{b_2}^{\bb_2}(\th_b)
     \; Z_{\ab_2}(\thb_a)Z_{\bb_2}(\bar{\th_b}) Z_w(0). \nonumber
\eea
As in the derivation of the Yang-Baxter equation, factorization now implies
that the order in which the particles scatter is  irrelevant too.
If we go through the same steps, but scatter particle $b$ first from the wall,
w
   e derive the following identity:
\beq
\label{YB1}
   ^-S_{ab}^{a_1b_1}(\th_{ab})W_{a_1}^{\ab_1}(\th_a)
    ^0S_{b_1\ab_1}^{b_2\ab_2}(\tilde{\th}_{b_1 a_1}) W_{b_2}^{\bb_2}(\th_b)=
  W_a^{\bb}(\th_b) ^0S_{a\bb}^{a_1\bb_1}(\tilde{\th}_{ab})
W_{a_1}^{\ab_1}(\th_a
   )
   ^+S_{\bb_1\ab_1}^{\bb_2\ab_2}(\th_{ab}).
\eeq
Diagramatically this corresponds to the equation in figure 1.

The presence of the wall breaks parity invariance, which - if true - would
deman
   d
$S_{ab}^{cd}(\th)=S_{ba}^{dc}(\th)$. But restrictions of this kind can be
genera
   ted, following the argumentation originally due to Cherednik
\cite{Ch1,Ch2}.
In the limit, when  the rapidity of one of the particles vanishes, it is
impossi
   ble to decide, whether it has or has not hit the wall before scattering off
a
   nother particle. This imposes the additional conditions:
\bdm
 W_{a}^{\ab}(0) ^+S_{ b \ab}^{ b_1 \ab_1}(\th)=
 ^0\!S_{ab}^{ a_1 b_1}(\th) W_{a_1}^{\ab_1}(0),
\edm
\beq
\label{YB2}
 W_{a}^{\ab}(0) ^0S_{b\ab}^{b_1 \ab_1}(\th)=
 ^-\!S_{ab}^{a_1 b_1}(\th) W_{a_1}^{\ab_1}(0).
\eeq
To complete the scheme, we still have to consider $3$-particle scattering.
Howev
   er if equ.(\ref{YB1}) is satisfied, we can always arrange rapidities, such
th
   at all particles scatter against each other, before ( or after ) they hit
the
    wall. Therefore  factorization requires $^{\pm}S(\th)$  to satisfy in
additi
   on the usual Yang-Baxter equations
\beq
^{\pm}S_{\ab \bb}^{\ab' \bb'}(\th_{ab})\; ^{\pm}S_{a \cb}^{\ab \cb' }(\th_{ac})
   \; ^{\pm}S_{bc }^{\bb \cb}(\th_{bc}) \;  =  ^{\pm}S_{\bb \cb }^{\bb' \cb'}
 (\th_{bc}) \;  ^{\pm}S_{\ab c}^{\ab' \cb }(\th_{ac}) \; ^{\pm}S_{ab}^{\ab \bb}
(\th_{ab})\;  \;\; . \label{eq: yangbax}
\eeq
These equations are sufficient to determine the $S$- and $W$-matrices, unless
th
   ey are diagonal. In this case (\ref{YB1}) are trivially satisfied and we
requ
   ire more information to determine them. Once an S-matrix posses a pole due
to the propagation of a bound state particle, one can formulate the so-called

\section{Bootstrap equations}
For simplicity we will in the sequel consider only diagonal $S,W$-matrices:
\beq
   W_{a}^{b}(\th)=\delta_{a}^{b} W_a(\th)
\eeq
and similarly for the $S$-matrices.
 In this case equs.(\ref{YB1}) and (\ref{YB2}) are satisfied, if
\beq
\label{diagS}
^0\!S_{ba}(\th)=^0\!S_{ab}(\th)=^-\!S_{ab}(\th)=^+\!S_{ba}(\th)
=^-\!S_{ba}(\th)=^+\!S_{ab}(\th).
 \eeq
Here we used unitarity equ.(\ref{unitarity}), which implies $W(0)^2=1$.
Equ.(\ref{diagS}) includes constraints usually coming from parity invariance.
As a result of this equation we shall not distinguish anymore in the following
between $^- S, ^0 S, ^+ S$ and solely refer to them as S.

When particle $c$ is a bound state of particles $a$ and $b$ one assumes in
addition to the Zamolodchikov algebra the validity of an operator product
expansion involving the operators representing those particles
\beq
Z_a\left(\th + i \eta_{ac}^b + \frac{i \varepsilon}{2} \right) \;\;
Z_b\left(\th - i \eta_{bc}^a - \frac{i \varepsilon}{2} \right) =
\frac{i \Gamma_{ab}^c Z_c(\th) }{\varepsilon}  \label{eq: OPE} \;\;  ,
\eeq
where $\Gamma_{ij}^{k}$ denotes the three particle vertex on mass-shell and
$\eta_{ab}^c$ are the so-called fusing angles.
Then multiplying this equation by $Z_d(\th_d)$, using equation (\ref{Zamalga})
and taking the limit $\varepsilon \rightarrow 0$ leads to a nontrivial
consistency condition,  which is known as the  {\em bootstrap} equation
\cite{ZZ
   }
\beq
S_{dc}(\th) = S_{da}( \th - i \eta_{ac}^b) \; S_{db}( \th + i \eta_{bc}^a) \; .
\label{eq: boots}
\eeq
It states that scattering particle $d$ against $c$ is equivalent to scatter $d$
   against the bound state $a+b$. Evidently there has to be an equation of this
kind in the presence of reflecting boundaries.
Thus let us scatter particles $a,b$ and $d$ with rapidities $\th_0+ i
\eta_{ac}^
   b
 + \frac{i \varepsilon}{2}  ,\th_0 - i \eta_{bc}^a - \frac{i \varepsilon}{2},
      \th_d >0$. We obtain by the same procedure as in the previous subsection
\bea
Z_a \!\!\!\!\! & & \!\!\!\!\!\left(\th_0 + i\eta_{ac}^b + \frac{i
\varepsilon}{2
   }
\right) Z_b\left(\th_0 - i \eta_{bc}^a - \frac{i \varepsilon}{2} \right)
 Z_d(\th_d)Z_w(0)
 =  S_{ab}\left(2 \th_{0}+  i \eta_{ac}^b -  i \eta_{bc}^a  \right) \nonumber\\
 & & S_{ad}\left(\th_{0d}+  i \eta_{ac}^b + \frac{i \varepsilon}{2} \right)
  S_{bd}\left(\th_{0d} -  i \eta_{bc}^a  - \frac{i \varepsilon}{2} \right)
  S_{ad}\left(\tilde{\th}_{0d}+  i \eta_{ac}^b + \frac{i \varepsilon}{2}
\right)
  \nonumber\\ & &
 S_{bd}\left(\tilde{\th}_{0d} - i \eta_{bc}^a  - \frac{i \varepsilon}{2}
\right)
  W_{a}\left(\th_a +  i \eta_{ac}^b + \frac{i \varepsilon}{2} \right)
  W_{b}\left(\th_b  - i \eta_{bc}^a - \frac{i \varepsilon}{2} \right)
  W_{d}(\th_d)   \nonumber  \\
& &Z_b\left(-\th_0 + i \eta_{bc}^a + \frac{i \varepsilon}{2} \right)
 Z_a\left(-\th_0 -  i \eta_{ac}^b - \frac{i \varepsilon}{2} \right)
 Z_{d}(- \th_{d})  Z_w(0) \;.  \nonumber
\eea
On the other hand multiplying equation (\ref{eq: OPE}) by $Z_d(\th_d)Z_w(0)$
and performing similar operations we derive after taking the limit
$\varepsilon \rightarrow 0$, the bootstrap equation in the presence of the wall
\bea
S_{dc}\!\!\!\!\! & & \!\!\!\!\!(\th_{0d})\; S_{dc}(\tilde{\th}_{0d})
W_c(\th_c)=
W_{a}(\th_a + i \eta_{ac}^b )  W_{b}(\th_b  - i \eta_{bc}^a )\label{bswall}\\
& &   \!\!\!\!\!\! S_{ab}(2 \th_{0}+  i \eta_{ac}^b -  i \eta_{bc}^a )  S_{ad}
(\th_{0d}+  i \eta_{ac}^b ) S_{bd}(\th_{0d} -  i \eta_{bc}^a )
S_{ad}(\tilde{\th}_{0d}+  i \eta_{ac}^b )S_{bd}(\tilde{\th}_{0d} - i
\eta_{bc}^a
     ). \nonumber
\eea
Diagramatically we depict this situation in figure 2. Since the
scattering matrix satisfies the conventional bootstrap equation
(\ref{eq: boots}), our equation for $W$ reduces to
\beq
 W_c(\th)=  W_{a}(\th + i \eta_{ac}^b) \; W_{b}(\th  - i \eta_{bc}^a )\;
 S_{ab}(2 \th +  i \eta_{ac}^b -  i \eta_{bc}^a ) \label{eq: inhomboot}
\eeq
which we call, due to the presence of the factor $ S_{ab}(2 \th_{0} +  i
\eta_{a
   c}^b -  i \eta_{bc}^a ) $, an {\em inhomogeneous} bootstrap equation.

The equations (\ref{YB1}) and (\ref{bswall}) also solve the analogous problem
of
    factorized scattering in the presence of two walls, since  the two walls do
   not interfere with each other.

%
\section{The $W$-matrix}

In this section, we shall discuss the solutions of the coupled equs.
(\ref{eq:  inhomboot}). The two particle scattering matrices to be used in
this paper always factorize into the form $S( \theta ) = \prod\limits_x
\{ x \}_{\theta} $. Adopting our notation from \cite{FO}, each of this block
reads
\beq
\label{eq: sblock}
\{ x\}_{\theta} := \frac{ [x]_{\theta} } { [-x]_{\theta} },
\eeq
with
\beq
 [x]_{\theta} := \frac{ \langle x + 1 \rangle_{\theta} \langle x - 1
\rangle_{\theta}  } { \langle x + 1 -B \rangle_{\theta} \langle x - 1 + B
\rangle_{\theta}  }
\eeq
and
\beq
\langle x \rangle_{\theta}  := \sinh \frac{1}{2} \left( \th + \frac{ i \pi x}
{h}  \right) \,\,  .
\eeq
 B is a function, which takes its values between 0 and 2 containing the
 dependence on the coupling constant $\beta$ of the Lagrangian field theory.
 h denotes the Coxeter number of the underlying Lie algebra of the theory. The
S-matrices possess furthermore the property to be invariant under $B
\rightarrow
    2 -B$, that is under an interchange of the strong and weak coupling regime.

Alternatively each block is equivalent to the following integral representation
\beq
\{ x \}_{\theta} = \exp \left( \int_{0}^{\infty} \frac{ dt} {t \sinh t}
\; f_{x,B}(t) \;\; \sinh \frac{ \theta t} {i \pi} \right) \label{eq: intrep}
\eeq
where
\beq
f_{x,B}(t) = 8 \sinh \frac{t B}{2 h} \sinh \frac{t}{h} \left( 1 - \frac{B}{2}
\right) \sinh t  \left( 1 - \frac{x}{h} \right) .
\eeq
Whilst (\ref{eq: sblock}) nicely exhibits the polestructure of the S-matrix,
equation (\ref{eq: intrep}) is sometimes more useful for explicit evaluations
and we shall require this form below.
\par
We might now expect that the $W$-matrix factorizes in an analogous fashion
into blocks as the S-matrix. Indeed we find a one-to-one correspondence between
   the blocks of the $W$- and $S$-matrix:
\beq
W( \theta ) \; =  \; \prod_x {\cal W}_x ( \theta ) \;\; . \label{eq: wfac}
\eeq
Similar as the S-matrix, the $W$-matrix factorizes further into subblocks
\beq
{\cal W}_x ( \theta ) \; = \; \frac{ w_{1-x} ( \theta)  w_{-1-x}(\theta) } {
w_{
   1-x - B} ( \theta)  w_{-1-x + B} ( \theta) } \;\; .\label{eq: subblock}
\eeq
As demanded by the unitarity (\ref{unitarity}) of the $W$-matrix we have
\beq
\label{eq: unit}
 w_x ( \theta ) \;\;  w_x (- \theta ) \; = \; 1.
\eeq
Furthermore we shall verify the relations
\bea
 w_{x-2h} ( \theta ) \;  w_{-x} ( \theta ) &=& 1 \\
 w_x ( 0 ) =  w_{-h} ( \theta )  &=& 1 \\
 w_x \left( \theta + \frac{i y \pi}{2h}\right)\;
 w_x \left( \theta - \frac{i y \pi}{2h}\right)  &=&
 w_{x+y} ( \theta )  \;  w_{x-y} ( \theta )  \\ \label{eq:ipi}
 w_{x} ( \theta + i \pi ) &=& \eta_x(\theta)  \; w_{x} (\th) \label{eq: etash}
\eea
where the function $\eta_x(\theta)$ satisfies individually the homogeneous
 bootstrap equation
\beq
\eta_x( \theta +i \eta_{ac}^b ) \; \eta_x( \th - i \eta_{bc}^a ) \; =\;
\eta_x(\theta) \; .
\eeq
Notice that $\eta_x(\theta)$  does not contain any poles in the physical sheet
$ 0 < \th < i \pi$. All blocks converge to one in the asymptotic limit $\th
\rightarrow 0$, resulting
from
\beq
\lim_{\th \rightarrow \infty}  w_x(\th) = 1  \;\; . \label{eq: asymp}
\eeq
It turns out that the function $w_x( \th )$  posses neither poles nor zeros
in the physical strip, such that no particle creation and absorption takes
place
in the wall.  The absence of the poles and zeroes was expected from the
assumpti
   on
that the scattering off the wall takes place in an elastic fashion.

 We shall now compute some explicit examples of the $W$-matrix, starting with
th
   e
%
\subsection{The Bullough-Dodd model}
The BD-model \cite{DB} ( $A_2^{(2)}$-affine Toda theory ) represents an
integrab
   le
quantum field theory involving one type of scalar field only, which satisfies
a relativistically invariant equation in two dimensions. The model is ideal
to illustrate the general principles presented in the previous sections, since
the particle, say A, emerges as a bound state of itself, i.e.
$A+A \rightarrow A$ is possible. Its classical Lagrangian is obtainable from
a folding \cite{SH,OT} of the $D_4^{(1)}$-affine Toda theory, where the three
ro
   ots
corresponding  to the degenerate particles and the negative of the highest
root are identified. The resulting Dynkin diagram is the simplest example of a
non-simply laced one, containing the root $\alpha$, which is related to the
scalar field, whose square length equals two and the root $\alpha_0 = - 2
\alpha$, whose square length is consequently eight. Then its classical
Lagrangian density reads
\beq
{\cal L}\,=\,\frac{1}{2}\partial_{\mu}\phi\partial^{\mu} \phi -
\frac{m^2}{\beta ^2}\left(2e^{ \beta \phi}+e^{-2 \beta\phi}\right)\,\, ,
\label{BuD}
\eeq
where $m$ denotes the bare mass and $\beta$ the coupling constant, which we
assume to be real to avoid the presence of solitons.
For more details on the model we refer to \cite{FMS2} and the references
therein
   , but here we mainly want state the main properties we are going to employ.
The scattering matrix can be obtained by the above folding procedure and it
turns out to be \cite{AFZ}
\beq
S^{BD}( \theta) \; = \; \{ 1 \}_{\theta} \; \{ 2 \}_{\theta} \,\,  .
\eeq
The Coxeter number of  the BD-model equals three.
\par
The scattering matrix satisfies the homogeneous bootstrap equation in the form
\beq
S^{BD}(\theta) \,= \,S^{BD}\left(\theta + \omega \right)  \;\; S^{BD}
\left(\theta - \omega \right) \;\; ,
\eeq
with $\omega = \frac{i \pi }{3}$ and consequently the inhomogeneous bootstrap
eq
   uation acquires the form
\beq
W(\th) \; = \; W\left(\th + \omega \right) \;  W\left(\th - \omega \right) \;
S^
   {BD}(2 \theta)  \;\; .
\eeq
Employing now Fourier transforms  after taking the logarithm to solve such
equations, we obtain
\beq
\label{WW}
W( \theta ) =  \exp \left( \;\;\int d \theta' \;
 G(\th - \th') \ln\,S(2 \theta') \right)  ,
\eeq
where the Green function $G$ is given by
\beq
G (\th)= \lim_{\eta\uparrow 1} \frac{1}{ \omega \sqrt{3}}\frac
{ \sinh \left( \frac{2 \pi}{3 \omega} ( \theta)\eta \right) }
{ \sinh \left( \frac{\pi}{3} ( \theta)\eta \right) } \;\;  .
\eeq
The introduction of the parameter $\eta$ is necessary to guarantee the
convergen
   ce of the Fourier transform. Employing now the
integral representation for the blocks of the S-matrix (\ref{eq: intrep}), we
are lead to a factorization of the form (\ref{eq: wfac}) and
(\ref{eq: subblock}), where each of the sublocks $w_x(\th)$ is given by the
integral representation
\beq
 w_x ( \theta ) = \exp  \left(\;\;\int_0^{\infty} \frac{ dt }{ t \sinh t}
\frac{ 2 \sinh \left( 1 + \frac{x}{h} \right)t \sinh \frac{2 \theta t}{i \pi} }
{ 1 - 2 \cosh\frac{2 t \omega}{\pi}  }  \right) \;\;  .
\eeq
Solving the integral we obtain
\beq
 w_x ( \theta ) \; = \; \prod_{l=0}^{\infty} \left( \frac{
\Gamma\left(1+(l+1)\frac{\omega}{\pi} + \frac{x}{2h} +\frac{i
\theta}{\pi}\right
   )
\Gamma\left((l+1)\frac{\omega}{\pi} - \frac{x}{2h} -\frac{i
\theta}{\pi}\right)}
   {
\Gamma\left((l+1)\frac{\omega}{\pi} - \frac{x}{2h} +\frac{i \theta}{\pi}\right)
\Gamma\left(1+(l+1)\frac{\omega}{\pi} + \frac{x}{2h}-\frac{i
\theta}{\pi}\right)
   }
\right)^{ \frac{\sin\left( (l + 1) \omega \right)}{\sin \omega }}
\;\;  . \label{eq: infprodgam}
\eeq
This equation exhibits nicely the pole structure of $  w_x ( \theta ) $,
and therefore $W(\th)$, and can be used to prove the functional identities
(\ref{eq: unit}) - (\ref{eq: asymp}). The function $\eta(\th)$, which results
as a shift of $i \pi$ in equation (\ref{eq: etash}) takes on the form
\beq
 \eta_x ( \theta ) \; = \; \prod_{l=0}^{\infty} \left( \frac{
\left((l+1)\frac{\omega}{\pi} - \frac{x}{2h} -\frac{i \theta}{\pi}\right)
\left(-1 +(l+1)\frac{\omega}{\pi} + \frac{x}{2h} +\frac{i \theta}{\pi}\right)}{
\left(1+(l+1)\frac{\omega}{\pi} + \frac{x}{2h} -\frac{i \theta}{\pi}\right)
\left((l+1)\frac{\omega}{\pi} + \frac{x}{2h}+\frac{i \theta}{\pi}\right)}
\right)^{ \frac{\sin\left( (l + 1) \omega \right)}{\sin \omega}}
\eeq
and satisfies individually the homogeneous bootstrap equation
\beq
\eta_x( \th + i \omega) \; \eta_x( \th - i \omega) \; =\; \eta_x(\theta) \; .
\eeq
$\eta_x(\theta)$ does not posses any poles in the physical sheet.
Furthermore we derive from (\ref{eq: infprodgam}) the functional equation
\beq
 w_x ( \theta  + i \omega)  \;\;  w_x ( \theta - i \omega) =
 w_x ( \theta )  \frac{ \langle x \rangle_{ -2 \theta} } { \langle x
\rangle_{2 \theta} }
\eeq
from which we infer the crucial identity for the blocks of the
W-matrix
\beq
{\cal W}_x ( \theta)  =   {\cal W}_x ( \theta + i \omega) \; {\cal W}_x (
\theta
    - i \omega)
\{ x \}_{2 \theta} \;\;  .   \label{eq: wboot}
\eeq
This equation demonstrates explicitly that the factorization of $W$ occurs
in a one-to-one fashion with respect to the factorization of the S-matrix and
we finally obtain the solution for the $W$-matrix of the Bullough-Dodd model
\beq
W(\th) = {\cal W}_1(\th) {\cal W}_2(\th) .
\eeq
Notice that this function posses   neither poles nor zeros in the physical
sheet
   .
%
\subsection{The $A_2^{(1)}$-affine Toda theory}
The $A_2^{(1)}$-affine Toda theory is the most simple example of an affine Toda
   theory  \cite{MOP,BCDS} involving more than one particle. It contains two
particles of equal masses which are conjugate to each other, that is choosing
complex scalar fields we have $\phi_1^{*} = \phi_2$. Its classical Lagrangian
de
   nsity
\beq
{\cal L}\,=\,\frac{1}{2}\partial_{\mu}\phi\partial^{\mu} \phi -
\frac{m^2}{ \beta ^2}\left( e^{ \beta \sqrt{2} \phi_{2} } + e^{ \frac{\beta}{
 \sqrt{2}} \left( \sqrt{3} \;\phi_{1} - \phi_{2} \right) } +  e^{
-\frac{\beta}{
 \sqrt{2}} \left( \sqrt{3} \;\phi_{1} + \phi_{2} \right) } \right)
\eeq
possesses a $Z \!\!\!\! Z_3$-symmetry, in the sense that it is left invariant
under the transformation
\beq
\phi \rightarrow \left( \begin{array}{l}
e^{ \frac{2 \pi i}{3} n } \\
e^{ \frac{4 \pi i}{3} n }   \end{array} \right) \;  \phi
\qquad \qquad  \qquad  n=1,2,3 \;\; .
\eeq
 From the
three point couplings, which turn out to be $ C_{111}= - C_{222} = -i 3 \beta
m^2$ and $ C_{112} = C_{221} =0$ or the application of the fusing rule of
affine Toda theory \cite{PD,FLO,FO,HB} we obtain that the following processes
 are permitted
\bea
V_1 + V_1  &\rightarrow&  V_2 = V_{\bar{1}}  \label{eq: fusea} \\
V_2 + V_2  &\rightarrow&  V_1 = V_{\bar{2}}  \label{eq: fuseb}
\eea
where we denote the particles by $V_i$ with $i=1,2$.
The scattering matrices are given by
\bea
S_{11}(\th) = S_{22}(\th) &=&  \{ 1 \}_{\th}  \\
S_{12}(\th)  &=&  \{ 2 \}_{\th}  .
\eea
Here the blocks are again of the form (\ref{eq: sblock}) with $h=3$.
Where $S_{11}(\th)=S_{22}(\th)$ have poles at $\frac{2 \pi i}{3}$ describing
the processes (\ref{eq: fusea}) and (\ref{eq: fuseb}), whereas
$S_{12}(\th)$ has no poles in the physical sheet. Furthermore, the scattering
matrix satisfies the bootstrap equations
\bea
S_{l2}(\th) &=& S_{l1}(\th + i \omega) S_{l1}(\th - i \omega) \\
S_{l1}(\th) &=& S_{l2}(\th + i \omega) S_{l2}(\th - i \omega)
\eea
for $l=1,2$, $\omega = i \pi / 3$, together with the crossed versions of this.
S
   ince the scattering matrices involved satisfy the ordinary bootstrap
equation
   s,
the wall bootstrap equations reduce to
\bea
W_2( \th ) & = & W_1( \th + i \omega ) \; W_1( \th - i \omega )
\; S_{11} ( 2 \theta    )  \\
W_1( \th ) & = & W_2( \th + i \omega ) \; W_2( \th - i \omega )
\; S_{22} ( 2 \theta    )  \;\;   .
\eea
Together with equation (\ref{eq: wboot}) we notice that these equations are
solved by
\beq
W_1(\th) = W_2(\th) = {\cal W}_{1}(\th) \;\;  .
\eeq
Again $W(\th)$ introduces no poles nor zeros in the physical sheet. The fact
tha
   t $W_1(\th)$
equals $W_2(\th)$ is a consequence of the mass degeneracy of the theory, which
is reflected by the automorphism of the Dynkin diagram \cite{SH,OT}. The
folding

towards the Bullough-Dodd model introcuces an additional block in the W-matrix,
due to the identification of particle 1 and 2, in a similar fashion as for the
S-matrix.

\subsection{The $A_4^{(2)}$-affine Toda theory}
The $A_4^{(2)}$ affine Toda theory is the most simple example of an affine Toda
   theory, where the roots associated to the particles are connected by
more than one lace on the Dynkin diagram. It descibes two self-conjugate real
scalar fields whose classical mass ratio is $m_1^2 = (5-\sqrt{5})/(5+\sqrt{5})
m_2^2$. The roots involved in this theory might be constructed from a
$D_6^{(1)}$-affine Dynkin diagram, where the four roots forming the two
handles and the two roots which are connnected to the handles are identified.
The resulting roots are
\beq
\alpha_1 = -\frac{ 2 \sqrt{2}}{\sqrt{5}} \left( \sin \frac{ 2 \pi}{5} ,
\sin \frac{  \pi}{5}  \right)  \qquad \hbox{and} \qquad
\alpha_2 = \frac{ 4 \sqrt{2}}{\sqrt{5}} \left( \sin \frac{  \pi}{5}
\cos \frac{2 \pi}{5} ,\sin \frac{ 2 \pi}{5}\cos \frac{ \pi}{5}  \right)
\nonumber
\eeq
where the root corresponding to the affinisation $\alpha_0$ is the
negative of twice the sum of this two roots. In terms of this vectors the
Lagrangian density reads
\beq
{\cal L}\,=\,\frac{1}{2}\partial_{\mu}\phi\partial^{\mu} \phi -
\frac{m^2}{ \beta ^2}\left( e^{ \beta \alpha_0 \cdot \phi} + 2 e^{ \beta
\alpha_
   1
\cdot \phi } + 2 e^{ \beta \alpha_2 \cdot \phi} \right) \;\; ,
\eeq
from which we may compute the three point couplings $C_{111} = C_{222} = 0$
and $C_{221} \neq 0 $, $C_{112} \neq 0 $
such that the following fusing processes are possible
\bea
V_1 + V_1  &\rightarrow&  V_2   \label{fusinga} \\
V_2 + V_2  &\rightarrow&  V_1  \label{fusingb} \\
V_1 + V_2  &\rightarrow&  V_1 + V_2 \;\; .  \label{fusingc}
\eea
The corresponding scattering matrices turn out to be
\bea
S_{11}(\th)  &=&  \{ 1 \}_{\th}  \{ 4 \}_{\th} \\
S_{12}(\th)  &=&  \{ 2 \}_{\th}  \{ 3 \}_{\th}        \\
S_{22}(\th)  &=&  \{ 1 \}_{\th}  \{ 2 \}_{\th}  \{ 3 \}_{\th} \{ 4 \}_{\th} .
\eea
The Coxeter number $h$ is five in this case. Here $S_{11}(\th)$ has a single
pole  with negative residue at $\frac{3 \pi i}{5}$ and  one with positive
 residue at $\frac{2 \pi i}{5}$ describing the process (\ref{fusinga}).
$S_{12}(\th)$ has single poles with negative residues at $\frac{ \pi i}{5}$,
$\frac{2 \pi i}{5}$ and single poles with positive residue at
$\frac{3 \pi i}{5}$, $\frac{4 \pi i}{5}$ corresponding to (\ref{fusingc}).
$S_{2
   2}(\th)$ has a single pole with negative residue at $\frac{ \pi i}{5}$,
a single pole with positive residue at $\frac{4 \pi i}{5}$ related the the
fusing (\ref{fusingb}) and  double poles at $\frac{2 \pi i}{5}$
$\frac{3 \pi i}{5}$. The bootstrap equation are in this case
\bea
S_{l2}(\th) &=& S_{l1}(\th + i \omega) S_{l1}(\th - i \omega) \\
S_{l1}(\th) &=& S_{l2}(\th + i \omega) S_{l2}(\th - i \omega) \\
S_{l1}(\th) &=& S_{l1}(\th + 3 i \omega) S_{l2}(\th + i \omega) \\
S_{l2}(\th) &=& S_{l2}(\th - i \omega) S_{l1}(\th + 2 i \omega)
\eea
with $l=1,2$. Because of the previous equations, the wall bootstrap equations
reduce to
\bea
W_2( \th ) & = & W_1( \th + i \omega ) \; W_1( \th - i \omega )
\; S_{11}( 2 \theta    )  \\
W_1( \th ) & = & W_2( \th + i \omega ) \; W_2( \th - i \omega )
\; S_{22} ( 2 \theta    )  \\
W_1( \th ) & = & W_2( \th + i \omega ) \; W_1( \th - 3 i \omega )
\; S_{12} ( 2 \theta    )  \\
W_2( \th ) & = & W_1( \th + 2 i \omega ) \; W_2( \th - i \omega )
\; S_{12} ( 2 \theta    )  \;\;  .
\eea
Introducing the notation $W_{11}(\th) := W_1(\th)$ , $W_{22}(\th) := W_2(\th)$
and $W_{12}(\th) := W_2(\th) / W_1(\th) $ , these equations decouple and we
are left with the problem to solve
\beq
W_{ij}(\th) \; = \; W_{ij}(\th + i \omega) \; W_{ij}(\th - i \omega) \;
W_{ij}(\th + 3 i \omega) \;W_{ij}(\th - 3 i \omega) \; S_{ij}(2 \th) \; .
\label{eq: A22wboot}
\eeq
Again we utilise Fourier transforms  after taking the logarithm and obtain
\beq
\label{WWij}
W_{ij}( \theta ) =
\exp \left( \;\;\int d \theta' \;
 G(\th - \th') \ln\,S_{ij}(2 \theta') \right)  ,
\eeq
where the Green function $G(\th)$ in this case is given by
\beq
G(\th)=  \lim_{\eta\uparrow 1}
\frac{1}{ \omega \sinh{\frac{\pi \th \eta}{\omega}}}  \left( \frac
{ \sinh \left( \frac{ \pi}{3 \omega} ( \theta)\eta \right) }
{ \sin  \frac{2\pi}{3}  }  +
\frac{ \sinh \left( \frac{ 4\pi}{5 \omega} ( \theta)\eta \right) }
{ \sin  \omega + 3  \sin 3 \omega   }  +
\frac{ \sinh \left( \frac{ 2\pi}{5 \omega} ( \theta)\eta \right) }
{ \sin  3 \omega + 3  \sin 9 \omega   }  \right) \;\;   .
\eeq
Employing now again the integral representation for the S-matrix we obtain
the following integral representaion for the building blocks of the
W-matrix
\beq
  w_x ( \theta ) = \exp \left( I_{\frac{ 2 \pi}{3} }(\th) + I_{\omega} (\th) +
I_{ 3 \omega} (\th) \right)  \;\; ,
\eeq
whith
\beq
 I_a(\th) =  \frac{1}{5-6 \sin^2 a} \int_0^{\infty} \frac{ dt }{ t \sinh t}
\frac{ 2 \sinh \left( 1 + \frac{x}{h} \right)t \sinh \frac{2 \theta t}{i \pi} }
{ \cos a - 2 \cosh\frac{2 t \omega}{\pi}  }   \;\;  .
\eeq
Solving the integral gives
\beq
 w_x ( \theta ) \; = \; \prod_{l=0}^{\infty} \left( \frac{
\Gamma\left(1+(l+1)\frac{\omega}{\pi} + \frac{x}{2h} +\frac{i
\theta}{\pi}\right
   )
\Gamma\left((l+1)\frac{\omega}{\pi} - \frac{x}{2h} -\frac{i
\theta}{\pi}\right)}
   {
\Gamma\left((l+1)\frac{\omega}{\pi} - \frac{x}{2h} +\frac{i \theta}{\pi}\right)
\Gamma\left(1+(l+1)\frac{\omega}{\pi} + \frac{x}{2h}-\frac{i
\theta}{\pi}\right)
   }
\right)^{P_l}
\;\;  .
\eeq
with
\beq
P_l = \frac{ \sin \left( (l+1) \frac{2 \pi}{3} \right)}{\sin \left(
\frac{\pi}{3} \right) }+ \frac{ \sin \left( (l+1) \frac{\pi}{5} \right)}
{2 \sin \left( \frac{\pi}{5} \right) \left( 5 - 6 \sin^2 \left( \frac{\pi}{5}
\right)  \right)} +  \frac{ \sin \left( (l+1) \frac{3\pi}{5} \right)}
{2 \sin \left( \frac{3\pi}{5} \right) \left( 5 - 6 \sin^2 \left( \frac{3\pi}{5}
\right)  \right)}
\eeq
Again this equation is useful to extract the polestructure and to prove the
identities (\ref{eq: unit}) - (\ref{eq: asymp}). The function $\eta_x(\th)$ is
n
   ow given by
\beq
 \eta_x ( \theta ) \; = \; \prod_{l=0}^{\infty} \left( \frac{
\left((l+1)\frac{\omega}{\pi} - \frac{x}{2h} -\frac{i \theta}{\pi}\right)
\left(-1 +(l+1)\frac{\omega}{\pi} + \frac{x}{2h} +\frac{i \theta}{\pi}\right)}{
\left(1+(l+1)\frac{\omega}{\pi} + \frac{x}{2h} -\frac{i \theta}{\pi}\right)
\left((l+1)\frac{\omega}{\pi} + \frac{x}{2h}+\frac{i \theta}{\pi}\right)}
\right)^{ P_{l}} \;\;  ,
\eeq
satisfying the homogeneous bootstrap equation and having no poles in the
physica
   l sheet. Further we derive the relation
\beq
 w_x ( \theta + i \omega) \;\; w_x  ( \theta - i \omega)   w_x ( \theta + 3 i
\omega)  w_x ( \theta - 3 i \omega)  =  \frac{ \langle x \rangle_{ -2 \theta} }
 { \langle x  \rangle_{2 \theta} }
\eeq
from which we deduce
\beq
{\cal W}_{{ij}_{x}} ( \theta ) = {\cal W}_{{ij}_{x}} ( \theta + i \omega) {\cal
W}_{{ij}_{x}} ( \theta - i \omega)  {\cal W}_{{ij}_{x}} ( \theta + 3 i \omega)
 {\cal W}_{{ij}_{x}}
 ( \theta - 3 i \omega)  \{x \}_{2 \th}.
\eeq
Comparision with equation (\ref{eq: A22wboot}) now demonstrates that the
$W$-matrix again factorizes in a one-to-one fashion with respect to the
scattering matrix and we finally obtain the $W$-matrix for the
$A_2^{(4)}$-affine Toda theory
\bea
W_1(\th) &= & {\cal  W}_1 {\cal  W}_4   \\
W_2(\th) &= & {\cal  W}_1 {\cal W}_2 {\cal  W}_3 {\cal  W}_4
\eea
{}From the property of the function $w_x(\th)$ we note again that the
physical sheet is free of singularities.
\section{Conclusions}
We have demonstrated how to formulate factorization equations and in particular
the inhomogeneous bootstrap equations by employing an extended version of
Zamolodchikov's algebra. Whereas in the absence of reflecting
boundaries such equations could be utilised to construct the two particle
scattering matrix, now they are sufficient to determine the $W$-matrix, which
encodes the scattering of a particle off the wall. For all cases investigated
$W(\th)$ does  posses neither poles nor zeros in the physical sheet, such that
the wall does not create or absorb any particles. This feature is made
transparent by expressing $W(\th)$ as infinite products of $\Gamma$ functions.
A
   ctually the structure is very similar to the one found for the minimal two
pa
   rticle form factors $F(\th)$ \cite{FMS,FMS2}.


\vfill \break
\begin{figure}
\setlength{\unitlength}{0.0125in}
\begin{picture}(40,0)(60,470)
\thicklines
\put(120,510){\line( 1,0){170}}
\put(340,510){\line(1,0){170}}
\put(120,500){\line(2,1){20}}
\put(130,500){\line(2,1){20}}
\put(140,500){\line(2,1){20}}
\put(150,500){\line(2,1){20}}
\put(160,500){\line(2,1){20}}
\put(170,500){\line(2,1){20}}
\put(180,500){\line(2,1){20}}
\put(190,500){\line(2,1){20}}
\put(200,500){\line(2,1){20}}
\put(210,500){\line(2,1){20}}
\put(220,500){\line(2,1){20}}
\put(230,500){\line(2,1){20}}
\put(240,500){\line(2,1){20}}
\put(250,500){\line(2,1){20}}
\put(260,500){\line(2,1){20}}
\put(270,500){\line(2,1){20}}
\put(340,500){\line(2,1){20}}
\put(350,500){\line(2,1){20}}
\put(360,500){\line(2,1){20}}
\put(370,500){\line(2,1){20}}
\put(380,500){\line(2,1){20}}
\put(390,500){\line(2,1){20}}
\put(400,500){\line(2,1){20}}
\put(410,500){\line(2,1){20}}
\put(420,500){\line(2,1){20}}
\put(430,500){\line(2,1){20}}
\put(440,500){\line(2,1){20}}
\put(450,500){\line(2,1){20}}
\put(460,500){\line(2,1){20}}
\put(470,500){\line(2,1){20}}
\put(480,500){\line(2,1){20}}
\put(490,500){\line(2,1){20}}
\put(310,540){$=$}
\put(220,510){\vector(1,3){25}}
\put(220,510){\line(-1,3){25}}
\put(410,510){\vector(1,3){25}}
\put(410,510){\line(-1,3){25}}
\put(170,510){\vector(2,1){120}}
\put(170,510){\line(-2,1){50}}
\put(460,510){\vector(2,1){50}}
\put(460,510){\line(-2,1){120}}
\put(125,535){$b$}
\put(136,514){$\theta_b$}
\put(187,575){$a$}
\put(390,575){$a$}
\put(350,568){$b$}
\end{picture}
 \caption{The factorization equation in the presence of a wall}
 \end{figure}
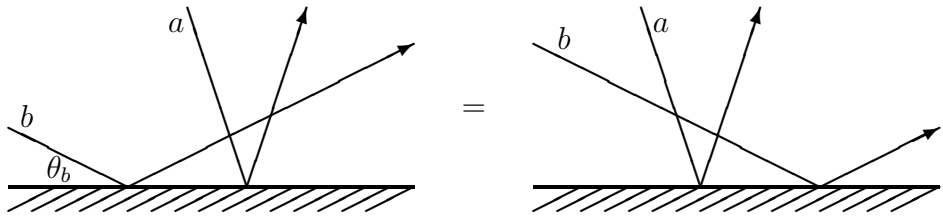
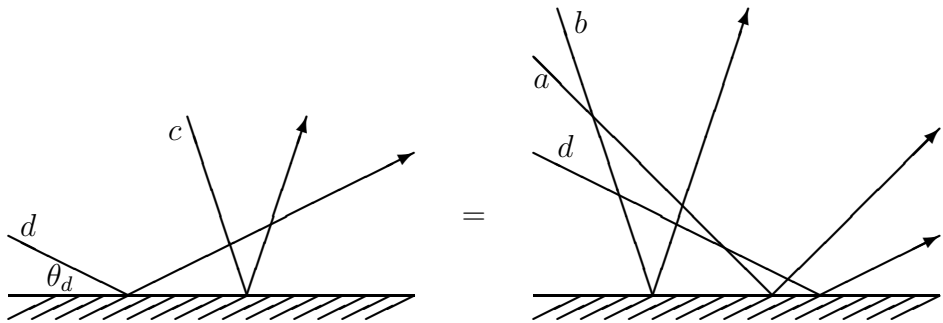
\begin{figure}
\setlength{\unitlength}{0.0125in}
\begin{picture}(40,90)(60,420)
\thicklines
\put(120,510){\line( 1,0){170}}
\put(340,510){\line(1,0){170}}
\put(120,500){\line(2,1){20}}
\put(130,500){\line(2,1){20}}
\put(140,500){\line(2,1){20}}
\put(150,500){\line(2,1){20}}
\put(160,500){\line(2,1){20}}
\put(170,500){\line(2,1){20}}
\put(180,500){\line(2,1){20}}
\put(190,500){\line(2,1){20}}
\put(200,500){\line(2,1){20}}
\put(210,500){\line(2,1){20}}
\put(220,500){\line(2,1){20}}
\put(230,500){\line(2,1){20}}
\put(240,500){\line(2,1){20}}
\put(250,500){\line(2,1){20}}
\put(260,500){\line(2,1){20}}
\put(270,500){\line(2,1){20}}
\put(340,500){\line(2,1){20}}
\put(350,500){\line(2,1){20}}
\put(360,500){\line(2,1){20}}
\put(370,500){\line(2,1){20}}
\put(380,500){\line(2,1){20}}
\put(390,500){\line(2,1){20}}
\put(400,500){\line(2,1){20}}
\put(410,500){\line(2,1){20}}
\put(420,500){\line(2,1){20}}
\put(430,500){\line(2,1){20}}
\put(440,500){\line(2,1){20}}
\put(450,500){\line(2,1){20}}
\put(460,500){\line(2,1){20}}
\put(470,500){\line(2,1){20}}
\put(480,500){\line(2,1){20}}
\put(490,500){\line(2,1){20}}
\put(310,540){$=$}
\put(220,510){\vector(1,3){25}}
\put(220,510){\line(-1,3){25}}
\put(170,510){\vector(2,1){120}}
\put(170,510){\line(-2,1){50}}
\put(125,535){$d$}
\put(136,514){$\theta_d$}
\put(187,575){$c$}
\put(460,510){\vector(2,1){50}}
\put(460,510){\line(-2,1){120}}
\put(440,510){\vector(1,1){70}}
\put(440,510){\line(-1,1){100}}
\put(390,510){\vector(1,3){40}}
\put(390,510){\line(-1,3){40}}
\put(350,568){$d$}
\put(340,597){$a$}
\put(357,620){$b$}
\end{picture}
 \caption{The inhomogeneous bootstrap equation}
 \end{figure}

\end{document}